\documentclass[%
 reprint,
superscriptaddress,
 amsmath,amssymb,
aps,
pra,
floatfix,
showkeys,
]{revtex4-2}

\usepackage{graphicx}
\usepackage{dcolumn}
\usepackage{bm}
\usepackage{hyperref}
\usepackage{amssymb} 

\usepackage[
scale=0.86, a4paper, marginratio={1:1, 1:1}, ignoreall,
]{geometry}

\begin{document}

\preprint{APS/123-QED}

\title{N\'eel domain wall as a tunable filter for optically excited magnetostatic waves}

\author{N.E. Khokhlov}
\email{n.e.khokhlov@mail.ioffe.ru}
\homepage{http://www.ioffe.ru/ferrolab/}
 \affiliation{Ioffe Institute, 26 Politekhnicheskaya, 194021 St. Petersburg, Russia}
\author{A.E. Khramova}%
 \affiliation{Ioffe Institute, 26 Politekhnicheskaya, 194021 St. Petersburg, Russia}
\affiliation{Faculty of Physics, M.V. Lomonosov Moscow State University, 119991 Moscow, Russia}
\affiliation{Russian Quantum Center, Skolkovo, 121205 Moscow, Russia}
\author{Ia.A. Filatov}
\affiliation{Ioffe Institute, 26 Politekhnicheskaya, 194021 St. Petersburg, Russia}%
\author{P.I. Gerevenkov}
\affiliation{Ioffe Institute, 26 Politekhnicheskaya, 194021 St. Petersburg, Russia}%
\author{B.A. Klinskaya}
\affiliation{Academic lyceum "Physical-Technical High School", 194021 St. Petersburg, Russia}%
\author{A.M. Kalashnikova}
\affiliation{Ioffe Institute, 26 Politekhnicheskaya, 194021 St. Petersburg, Russia}%

\date{\today}

\begin{abstract}

We present a concept of a tunable optical excitation of spin waves and filtering their spectra in a ferromagnetic film with 180$^{\circ}$ N\'eel domain wall.
We show by means of micromagnetic simulation that the fluence of the femtosecond laser pulse and its position with respect to the domain wall affect the frequencies of the excited spin waves, and the presence of the domain wall plays crucial role in control of the spin waves' spectrum.
The predicted effects are understood by analyzing the changes of the spin waves' dispersion under the impact of the laser pulse.
\end{abstract}

\maketitle


\section{Introduction}

In magnonics, spin waves (SW) are used to implement alternative methods of transferring information in magnetic nanostructures that can replace traditional transistor circuits \cite{Lenk-PhysRep2011, Nikitov:UFN2015,ChumakNPhys:2015, Mahmoud_JAP_2020_Intro_to_SW_computing}. 
Unlike electric charges, SW can propagate in materials even without free charge carriers \cite{Kajiwara_Nature2010:Transmission_SW_in_YIG, hou_spin_antiferromagnets_2019_NPGAsia}.
Thus, SW propagation is not associated with Joule losses which reduction is the challenging problem in traditional electronics. 
Different types of magnetic ordering support SW with frequencies in the range from GHz to THz \cite{wang2002spin, cramer_magnon_2017_NanoLett, rezende_AntiFerro_magnonics_2019_JAP, Lebrun_long-distance_spin-transport_2020_NatComm}, extending the operation rates of magnonics circuits.

Developing approaches for controlling SW is essential for bringing magnonics concepts to  applications.
Control of SW amplitude, phase, velocity, and propagation direction have been demonstrated by introducing various types of magnetic non-uniformity in the SW guiding media \cite{vasiliev2007spin, sadovnikov_magnonic_beamsplitter_2015_APL, StigloherPRL:2016, Hioki_bi-reflection_2020_CommunPhys}. 
Particularly, topological defects such as domain walls (DW) and skyrmions change amplitude and phase of SW passing through them \cite{hertel2004domain, buijnsters2016chirality, Chang_SciRep2018ferromagneticDW_filter, albisetti_synthetic_2020, hamalainen_control_SW_transm_2018_NatComm, Han_DWmotion_bySW_2009_APL, Dadoenkova_inelastic_SW_scattering_2019_PSSRRL, Yan_SWandDW_PRL_2011, Wang_DW_motion_via_SW_PRB_2012, Han_Mutualcontrol_Science_2019, Lan_SW_skyrmion_skew_PRB_2021}.
In parallel, tunable magnetic non-uniformities, e.g., those induced by illuminating a magnetic structure with light, are of particular interest since they allow creating reconfigurable magnonic elements \cite{VogelNPhys:2015, grundler_reconfigurable_2015, VogelSciRep:2018,Sadovnikov:PRB2019}.
Furthermore, changing magnetic properties of a medium locally by femtosecond laser pulses appears to be one of the efficient ways to generate propagating spin waves and to tune their characteristics \cite{Hioki_bi-reflection_2020_CommunPhys, satoh_2012directional_control_SW, JacklPRX:2017, Khokhlov_SWinFeGa:PRApplied2019, Au:2013_PRL_DirectExcitation, Muralidhar_Caustic_SW_beams_2021_PRL}. 

There is a broad range of mechanisms enabling manipulation of magnetic ordering by laser pulses \cite{KIMEL_Fundamentals_PhysReports_2021, Kirilyuk2010review, Walowski_Ultrafast_and_THz_JAP_2016}, including ultrafast demagnetization, inverse magneto-optical effects, excitation of coherent phonons, ultrafast change of magnetic anisotropy.
The last one is a versatile mechanism as the anisotropy could be of different nature: magnetocrystalline, shape-, strain-induced, etc.
Furthermore, laser-induced anisotropy changes as a triggering mechanism of magnetization dynamics can be realized in metals, semiconductors, and dielectrics \cite{Baranov:2019UFN} through ultrafast heating \cite{shelukhin_YIG_2018_PRB, Carpene_ultrafast_3D_anisotropy_change_PRB_2010, Bigot_Ultrafast_magnetization_anisotropy:ChemPhys_2005}, excitation of lattice distortions \cite{Kats_anisotropy_driven_phonos_PRB:2016}, photomagnetic effects \cite{stupakiewicz_ultrafastRecording_2017_Nature}, etc.
Therefore, it is appealing to realize a tunable source of SW by combining the advantages of ultrafast laser-induced excitation of SW with their control by local magnetic defects, such as DW.

In this Article, we present a micromagnetic study of magnetostatic spin waves (MSW) optically excited in the vicinity of a 180$^{\circ}$ N\'eel DW in a thin ferromagnetic film.
MSW are triggered by changes of the magnetic anisotropy of the film resulting from ultrafast laser-induced increase  of the lattice temperature occurring within a few picoseconds after the excitation \cite{shelukhin_YIG_2018_PRB, Carpene_ultrafast_3D_anisotropy_change_PRB_2010, Bigot_Ultrafast_magnetization_anisotropy:ChemPhys_2005}.
We show that the laser-excited area and the DW effectively form a tunable source of MSW.
Excitation of selected frequencies in the MSW spectrum is found to be controlled by the laser spot - DW distance, as well as by laser pulse fluence and laser spot width. 

\section{Model details}

\begin{figure}
\includegraphics[width=1 \linewidth]{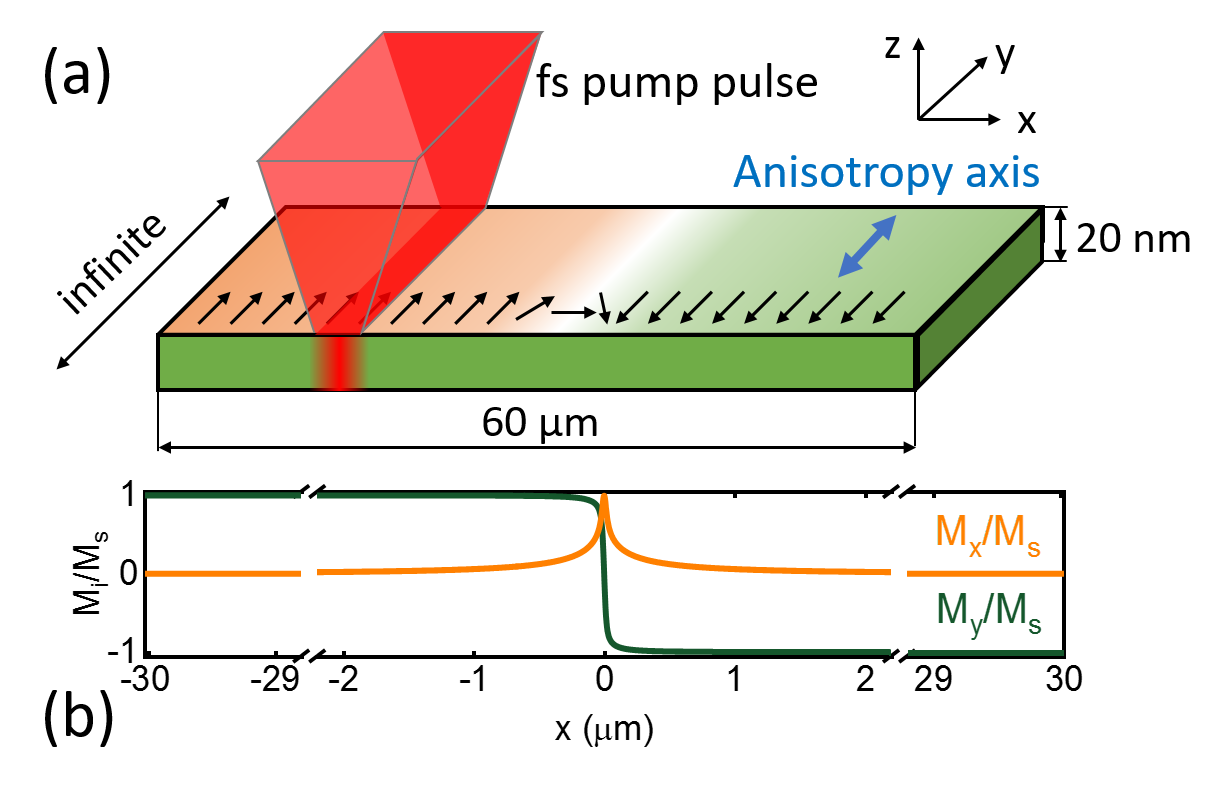}
\caption{\label{fig:Scheme} 
(a) Scheme of the ferromagnetic strip with 180$^{\circ}$ N\'eel domain wall at the center.
Small arrows indicate orientation of magnetization ${\bf M}$ at different $x$.
Bold double arrow indicates the anisotropy axis along $y$-axis.
(b) Calculated initial spatial distribution of the in-plane magnetization projections $M_x$ and $M_y$ across the strip; out-of-plane projection $M_z$ is zero.
}
\end{figure}

We use micromagnetic numerical calculations on a model system of a ferromagnetic strip.
The material parameters of the strip are: the saturation magnetization $M_s$~=~800\,kA/m, exchange stiffness $A$~=~1.3$\cdot$10$^{-11}$\,J/m, Gilbert damping constant $\alpha$=0.008, uniaxial anisotropy parameter $K_u$~=~5\,kJ/m$^3$.
The strip has a width of 60\,$\mu$m in the $x$-direction, infinite length in the $y$-direction, and a thickness of 20\,nm, as shown in Fig.\ref{fig:Scheme}(a).
Easy axis of magnetic anisotropy is along the $y$-axis.

The Object Oriented MicroMagnetic Framework OOMMF \cite{OOMMF} is utilized for solving the Landau–Lifshitz–Gilbert equation \cite{landau1935theory, Gilbert_theory_of_damping_IEEE_2004}:
\begin{equation}\label{Eq:LLG}
    \frac{d \bf M}{dt} = -  |\gamma|  {\bf M}\times{\bf H}_\mathrm{eff} + \frac{\alpha}{M_s}\left({\bf M}\times \frac{d \bf M}{dt}\right),
\end{equation}
where $\bf M$ is magnetization, $t$ is time, $\gamma = -2.211\times10^5\,mA^{-1}s^{-1}$ is the Gilbert gyromagnetic ratio, ${\bf H}_\mathrm{eff}$ is the effective field consisting of anisotropy, exchange and magnetostatic fields.
We consider the case when no external magnetic field is applied.
The solution of Eq.\,(\ref{Eq:LLG}) is the dynamics of ${\bf M}$  as a function of $t$ following the excitation with the laser pulse.
We use a time step of 100\,fs and a total time window of 4\,ns. 
A cell size of $4\times4\times20$\,nm$^3$ is chosen to be smaller than both the magnetostatic exchange length $\sqrt{2A/(\mu _0M_s^2)}$ and the magnetocrystalline exchange length $\sqrt{A/K_u}$  \cite{abo_definition_Lex_IEEE_2013}.
The infinite strip length along the $y$-axis is modeled with 1D periodic boundary conditions.
The calculated initial distribution of ${\bf M}$ is a two-domain state with the 180$^{\circ}$ N\'eel DW at the center of the strip, $x_{0} = 0$; ${\bf M}$ is oriented in the film plane (Fig.\ref{fig:Scheme}(b)).

We assume the impact of the optical laser pulse as a local relative reduction of the anisotropy parameter $\Delta K_u$ resulting  from  the  laser-induced heating.
As found in various experiments, reduction of the anisotropy parameter occurs  at a  picosecond  time  scale \cite{Carpene_ultrafast_3D_anisotropy_change_PRB_2010, shelukhin_YIG_2018_PRB}, and,  thus, can be approximated in our model by the instantaneous decrease of $K_u$.
Following  recovery  of $K_u$ to  its  equilibrium value occurs at time scales of the order of a few nanoseconds, and is neglected in our model.

We consider the pump spot having Gaussian profile along $x$-axis and elongated infinitely along $y$-axis to model the excitation of plane SW, similar to recent all-optical experiments \cite{Hioki_snells_2020_APL,Hioki_bi-reflection_2020_CommunPhys, Matsumoto_SW_trough_gap_PRB_2020}.
The resulting spatial-temporal profile of the anisotropy change is
\begin{equation}\label{eq:delta_Ku}
    K_u(x,t) = K_u \left(1 -\Delta K_u \exp\left[\frac{(x-x_p)^2}{2\sigma^2} \right] \Theta(t) \right),
\end{equation}
where $x_p$ is the position of the pump spot center, $\Theta(t)$ is a Heaviside function, $\sigma$ characterizes the width of the pump.
In the following, we refer to $\sigma$ as the pump width.
It is worth noting that the minimum accessible value of $\sigma$ is limited by the diffraction limit and is of a few hundreds  of  nanometers.
Thus, the range of wavenumbers for optically excited SW propagating laterally has the upper limit of $\sqrt{2}/\sigma$ \cite{KamimakiPRB:2017, Khokhlov_SWinFeGa:PRApplied2019}.
Thus, wavenumbers lesser than 10\,rad/$\mu$m are exited in optical experiments, what corresponds to the magnetostatic type of SW.
Nonetheless, exchange perpendicular standing SW are accessible by optical excitation in flat metal films \cite{Kamimaki2017ieee, vanKampen:PRL2002} and complex dielectric structures \cite{Chernov_Exchange_Spin_Waves_NanoLett2020}.
Such waves do not propagate laterally and are omitted from the consideration below.

The analysis of MSW properties is based on monitoring the temporal and spatial evolution of the out-of-plane component of the magnetization $M_z$ as laser-induced MSW are usually detected in all-optical pump-probe experiments using polar magneto-optical Kerr effect in reflection \cite{Au:2013_PRL_DirectExcitation, Khokhlov_SWinFeGa:PRApplied2019, Kamimaki2017ieee, KamimakiPRB:2017} or Faraday effect in transmission \cite{satoh_2012directional_control_SW, Hioki_bi-reflection_2020_CommunPhys, Hioki_snells_2020_APL, JacklPRX:2017, Matsumoto_SW_trough_gap_PRB_2020}.
Similar to all optical pump-probe experiments, we assume the detection of $M_z$ at a variable position $x_{probe}$.

\section{Results and discussion}

\subsection{Excitation of spin waves}

\begin{figure}
\includegraphics[width=1\linewidth]{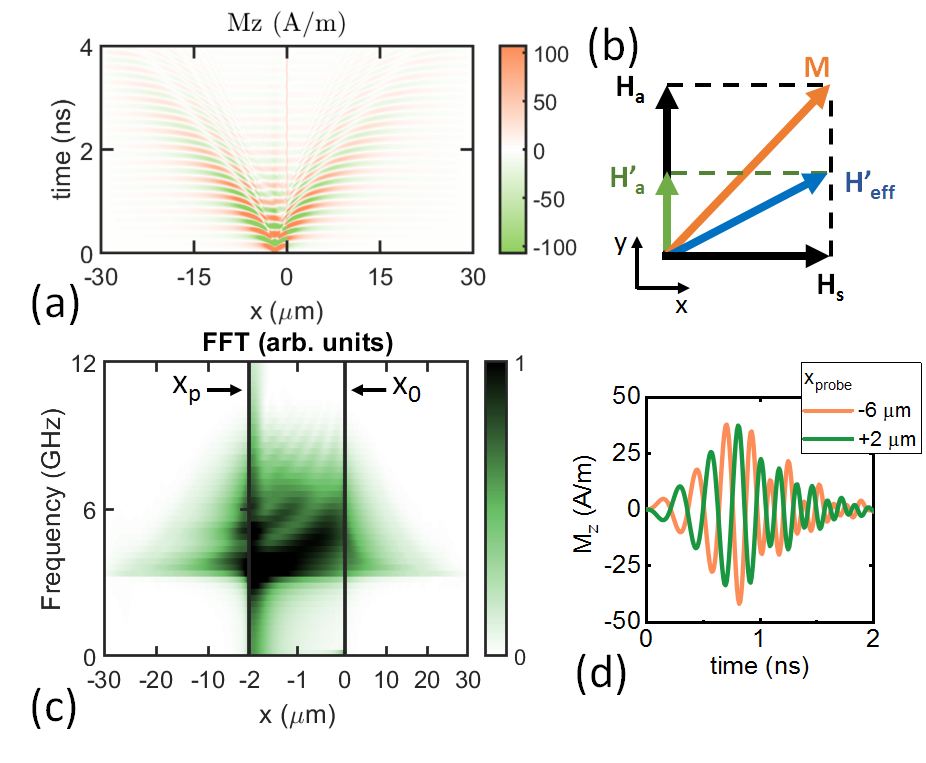}
\caption{\label{fig:time-space_map} 
(a) Spatial-temporal map $M_z(x,t)$.
(b) Scheme of reorientation of total effective field ${\bf H}_{eff}$ due to decreasing of anisotropy parameter.
${\bf H}_a'$ and ${\bf H}_{eff}'$ are the anisotropy and effective fields at $t=0+$, respectively.
(c) 1D FFT of $M_z(t)$ vs $x$-coordinate.
(d) $M_z(t)$ at \mbox{$x_{probe}=x_p\pm 4 \mu$m}.
Data in (a,c,d) are obtained for $x_p = -2\,\mu$m, $\sigma = 0.2\,\mu$m, $\Delta K_u = 0.2$.
}
\end{figure}

Figure \ref{fig:time-space_map}(a) shows the spatial-temporal distribution of $M_z$ for the pump pulse positioned at \mbox{$x_p=-2\,\mu$m}.
The calculations show the propagation of MSW packets from the pump spot in both directions along the $x$-axis.

The excitation of MSW is enabled by an abrupt reorientation of effective field ${\bf H}_{eff}$ caused by the pump, as shown schematically in Fig.\ref{fig:time-space_map}(b).
At equilibrium, ${\bf H}_{eff}$ is a sum of stray field ${\bf H}_s$ and anisotropy field ${\bf H}_a = (2K_u/M_s){\bf m}_y$, where ${\bf m}_y = (M_y/M_s){\bf e}_y$, ${\bf e}_y$ is the unit vector along the $y$-axis.
Non-zero component of ${\bf H}_s$ along $x$-axis is proportional to $M_x$.
Stray field appears because of the presence of the DW and decreases with the distance from the wall (Fig.\ref{fig:Scheme}(b)).
As $H_a$ decreases due to laser-induced change of the anisotropy parameter $\Delta K_u$, ${\bf H}_{eff}$ changes its magnitude and orientation to ${\bf H}_{eff}'$ at the time scale much shorter than the magnetization precession period.
This leads to a non-zero angle between ${\bf H}_{eff}'$ and ${\bf M}$ in the $xy$-plane at $t=0+$ with corresponding non-zero torque $\bf{T}$ acting on $\bf{M}$:
\begin{eqnarray}\label{eq:torque}
    {\bf T}(x) =  -|\gamma| {\bf M}(x)\times{\bf H}_{eff}'(x) = \nonumber \\
    = 2|\gamma| K_u \Delta K_u \frac{M_x(x)M_y(x)}{M_s^2}{\bf e}_z,
\end{eqnarray}
where ${\bf e}_z$ is the unit vector along the $z$-axis.
Non-zero \textbf{T} triggers the precession of \textbf{M} within the pump spot, which in turn launches the propagation of MSW outside the spot.

Equation (\ref{eq:torque}) shows that the described mechanism of MSW excitation requires presence of the DW.
Indeed, the absence of DW means the initial single domain state of the film with  ${\bf H}_{eff}$ and ${\bf M}$ aligned along $y-$axis and resulting $H_s = 0$ at any $x$.
Thus, the change of $K_u$ alters $H_a$, but does not affect the direction of ${\bf H}_{eff}$.
The presence of DW in the two-domain state provides non-zero $M_x$ (Fig.\ref{fig:Scheme}(b)) with the corresponding non-zero $x$-component of ${\bf H}_s$ in the vicinity of $x_0$.
Therefore, the presence of DW in our system enables MSW excitation via ultrafast changes of magnetic anisotropy even in zero applied magnetic field.

\subsection{Spectrum of MSW optically excited in the DW vicinity}

Here we turn to the analysis of the MSW features related to the DW presence.
The spatial-temporal maps $M_z(x,t)$ show the reflection and transmission of the MSW from/through DW (Fig.\ref{fig:time-space_map}(a)).
For detailed movie of the MSW propagation see Suppl. Mat. \cite{Supplementary}.
The reflection is due to nonuniform ${\bf H}_{eff}$ in the DW vicinity.
Indeed, the orientation of ${\bf H}_{eff}$ defines the MSW dispersion law $f(k)$ \cite{gurevich1996magnetization}, where $f$ and $k$ are MSW frequency and wavenumber, respectively.
Far from DW, ${\bf k}$ and ${\bf H}_{eff}$ are orthogonal and $f(k)$ corresponds to the surface mode of MSW \cite{damon_magnetostatic_1961}.
In the DW vicinity, ${\bf H}_{eff}$ acquires a projection on the $x$-axis resulting in corresponding changes of $f(k)$.
Thus, the DW works as a non-uniformity of the effective refractive index for MSW, and, as a consequence, a fraction of MSW packet is reflected from DW \cite{Chang_SciRep2018ferromagneticDW_filter}.
The reflected MSW interferes with the MSW propagating directly from the pump spot at $x<x_0$.
We note that the DW displacement resulting from the interaction with the MSW wavepacket is found to be of 4 nm only (1 cell of the mesh) and, thus is omitted from the consideration below.

To analyze in detail the MSW interference at various $x$, we performed one-dimensional fast Fourier transform \mbox{(1D FFT)} of the temporal signals at different $x$.
The resulting $x$-$f$ maps (Fig.\ref{fig:time-space_map}(c)) demonstrate an evolution of MSW packet's spectrum with $x$, possessing a fir-tree-shape.
In particular, there are multiple peaks in the FFT spectra at $x<x_0$ due to the interference.
The range $x>x_0$ does not reveal any pronounced interference pattern, and the MSW spectrum possesses a single broad maximum.
Thus, below, we focus our discussion on the properties of MSW at $x<x_0$.
However, we note that DW brings the phase shift of $\pi$ to the MSW packet at $x>x_0$ (Fig.\ref{fig:time-space_map}(d)).
The effect could be described as a switch of a spin angular momentum of magnons passing through the DW and was observed in resent experiments \cite{Han_Mutualcontrol_Science_2019}.

At $x<x_0$, we distinguish two ranges, $x_p<x<x_0$ and $x<x_p$, with different interference patterns and multipeak spectra.
The region $x_p<x<x_0$ can be seen as a resonator formed by DW and the area excited with the pump pulse.
Here DW works as a partially reflecting mirror for MSW, as described above, and the pump spot produces non-uniformity of $K_u$ and, thus, acts as a second mirror of the resonator.
Indeed, the changes of $K_u$ modify ${\bf H}_{eff}$ with corresponding variation of $f(k)$ inside the pump spot.
As a result the spot works as a potential gap for some MSW frequencies, as discussed below in Sec. D.
Detailed discussion about the traps of MSW induced by continuous wave lasers could be found elsewhere \cite{Kolokoltsev:APL2012, Busse:SciRep2015}.

In the region $x<x_p$, MSW exit the DW-pump resonator and possess spectrum with equidistant peaks.
As we show below, the properties of the resonator are defined by the pump parameters, and it enables controllable variations of MSW spectra.
Below, we focus on discussion of the region $x<x_p$ as the frequency composition of MSW packet does not varies with $x$ here.

\subsection{Effect of the pump position on the MSW spectrum}

\begin{figure}
\includegraphics[width=1\linewidth]{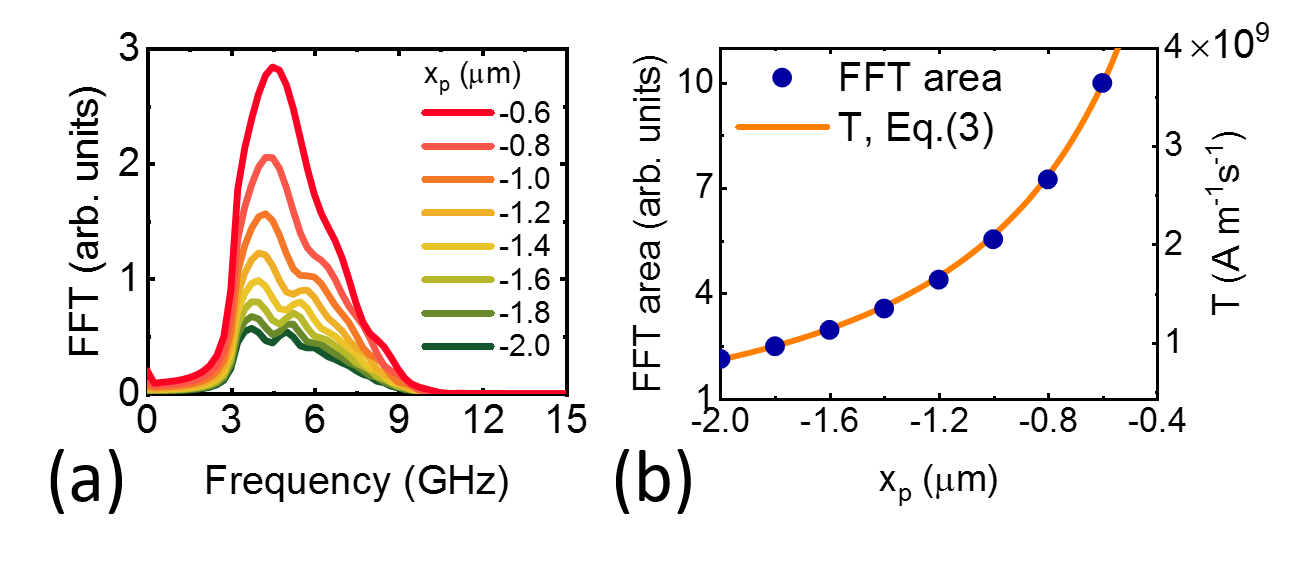}
\caption{\label{fig:FFT_vs_Xpump} 
(a) 1D FFT of $M_z(t)$ at different $x_p$ and $\Delta K_u = 0.1$, $\sigma = 0.2\,\mu$m, \mbox{$x_{probe} = x_p - 2\,\mu$m}.
(b) Area under the FFT curves (symbols) and torque $T(x)$ calculated using Eq.\,(\ref{eq:torque}) (line) versus $x_p$.
}
\end{figure}

Figure\,\ref{fig:FFT_vs_Xpump}(a) shows spectra of the MSW obtained at $x_{probe}=x_p-2\,\mu$m for different positions of the pump spot $x_p$.
As can be seen, the variation of $x_p$ affects the shape and amplitude of MSW spectrum at $x<x_p$.
This is a result of the changes of the DW-pump resonator length \mbox{$l_{res} = |x_0-x_p|$}, on the one hand, and the spatial variation of $H_s$ in the DW vicinity, on the other hand.

The pump position $x_p$ defines $l_{res}$, that, in turn, affects the resonance peaks in the spectrum of MSW and their spectral positions.
The number of peaks is also defined by $l_{res}$.
In particular, for larger $l_{res}$ more peaks with smaller distance between them occur within the MSW spectrum (Fig.\ref{fig:FFT_vs_Xpump}(a)).
Additionally, the choice of $x_p$ defines the amplitude of the excited MSW as follows.
$x_p$ defines the value of $M_x$ and $H_s$ inside the pump spot.
They are maximal in the DW vicinity, leading to a maximal laser-induced torque ${\bf T}$ (Eq.\,(\ref{eq:torque})) near the DW.
Thus, for smaller $l_{res}$, larger amplitude of MSW is observed.
To demonstrate it, we find  the  amplitude  of  the spectrally broad MSW packet as the area under the FFT curve and plot it as a function of $x_p$ in Fig.\,\ref{fig:FFT_vs_Xpump}(b).
As can be seen, the change of the MSW packet amplitude with $x_p$ follows the dependence $T(x)$ from Eq.(\ref{eq:torque}).

\subsection{Effect of pump fluence and width on the MSW spectrum}

\begin{figure}
\includegraphics[width=1\linewidth]{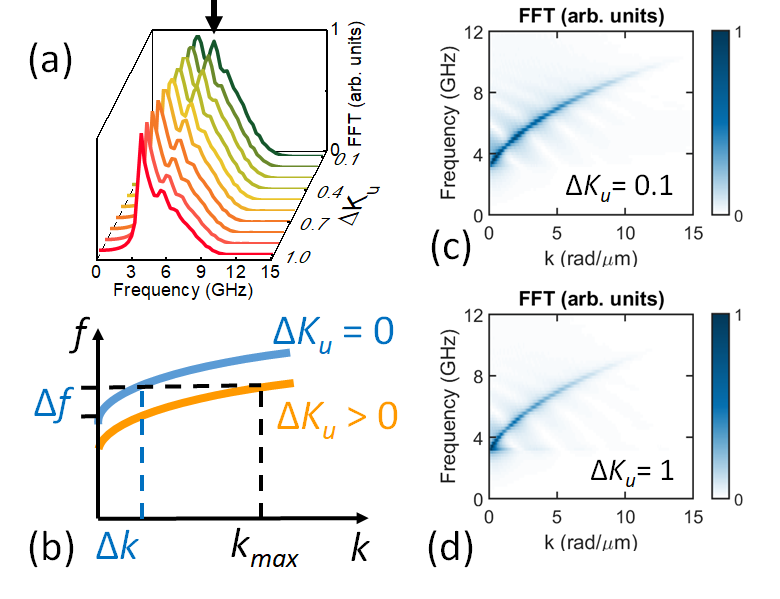}
\caption{\label{fig:FFT_vs_dK} 
(a) Normalized 1D FFT of $M_z(t)$ at different $\Delta K_u$ and \mbox{$x_p = -2\,\mu$m}, $\sigma = 0.2\,\mu$m, $x_{probe} = -4\,\mu$m.
Black arrow points at the additional resonance peak in the spectra.
(b) Scheme of dispersion shift with $\Delta K_u$.
(c,d) 2D FFT of $M_z(x,t)$ maps at $\Delta K_u = 0.1$ and $\Delta K_u = 1$, respectively.
}
\end{figure}

To simulate the effect of the optical pump fluence variation on the MSW properties, we performed calculations for different values of $\Delta K_u$ at constant $\sigma$.
As a result, the change of $\Delta K_u$ leads to the change of MSW spectrum.
Particularly, the increase of $\Delta K_u$ narrows the spectrum width and decreases the relative amplitude of high-frequency resonance peaks (Fig.\ref{fig:FFT_vs_dK}(a)).

The effect of pump fluence on MSW spectra could be described qualitatively in terms of $f(k)$ shift due to $\Delta K_u$ (Fig.\ref{fig:FFT_vs_dK}(b)).
The spectrum of the excited MSW packet is limited by two factors -- the time scale $\tau_K$ of $K_u$ decrease, and the range of excited MSW wavenumbers $k$.
Typical $\tau_K$ is within a few hundreds of fs in the experiments, and $1/\tau_K$ is within the THz range, which is substantially larger than MSW frequency in ferromagnets.
The maximum value of excited MSW wavenumbers $k_{max} = \sqrt{2}/\sigma$ is defined by the width of the spatial Fourier transform of the pump spot \cite{satoh_2012directional_control_SW, Khokhlov_SWinFeGa:PRApplied2019, KamimakiPRB:2017}.
Thus, the spectral range of the excited MSW spans from $f_{FMR}=f(0)$ to $f(k_{max})$ (Fig.\ref{fig:FFT_vs_dK}(b)).
The decrease of $K_u$ leads to a negative frequency shift of $f(k)$ inside the pump spot due to decrease of $H_{eff}$ (lower curve in Fig.\ref{fig:FFT_vs_dK}(b)).
Thus, a smaller spectral width of the MSW $\Delta f$ is observed outside the laser spot as there is a cut-on frequency $f_{FMR}$ (upper curve in Fig.\ref{fig:FFT_vs_dK}(b)).
The increase of $\Delta K_u$ leads to the smaller range of $\Delta f$ for MSW propagating outside the spot.

The effect of the MSW spectrum narrowing is seen in the calculation results (Fig.\ref{fig:FFT_vs_dK}(a)).
Notably, the ratio between the amplitudes of resonance peaks changes with $\Delta K_u$.
The increase of $\Delta K_u$ leads to narrowing of the range of the MSW wavenumbers $\Delta k$ detected outside of the pump spot, as follows from the $f(k)$ scheme (Fig.\ref{fig:FFT_vs_dK}(b)) and verified by reconstruction of $f(k)$ from the calculated $x$-$t$ maps with the 2D FFT (Fig.\ref{fig:FFT_vs_dK}(c,d)).

The dispersion scheme (Fig.\ref{fig:FFT_vs_dK}(b)) explains also the  spatial decay of the low-frequency part of the MSW spectrum evident in Fig.\ref{fig:time-space_map}(c).
Outside of the pump spot, only the high-frequency part of MSW packet can propagate as the lower frequencies are forbidden at $\Delta K_u=0$ (upper curve in Fig.\ref{fig:FFT_vs_dK}(b)).
As a result, the magnetization precession with $f < f_{FMR}$ is observed only within the laser-excited spot, which is seen as a "trunk" of the "tree" at $x = x_p$ on the $x$-$f$ map (Fig.\ref{fig:time-space_map}(c)).
The shape of "coma" is related to the stronger spatial decay of the high frequency part of the MSW packet upon propagation.
Indeed, the propagation length $L_{pr}$ of the MSW with certain $f$ is determined by its lifetime $\tau = (2\pi f \alpha)^{-1}$ and phase velocity $v_{ph}=2\pi f [k(f)]^{-1}$ as
\begin{equation}
    L_{pr}(f) = \tau v_{ph} = \frac{1}{\alpha k(f)}.
\end{equation}
Thus, the positive slope of $f(k)$ leads to faster spatial decay of MSW with higher $f$.

\begin{figure}
\includegraphics[width=1\linewidth]{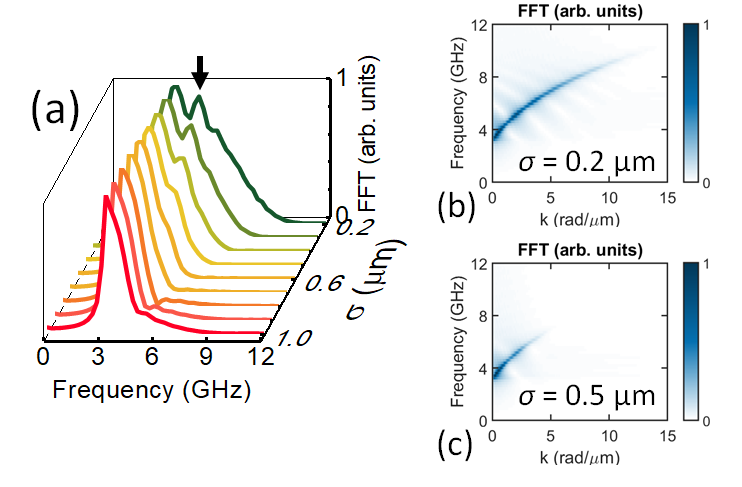}
\caption{\label{fig:FFT_vs_Sigma} 
(a) Normalized 1D FFT of $M_z(t)$ at different $\sigma$ and $x_p = -2\,\mu$m, $\Delta K_u = 0.2$, $x_{probe} = -4\,\mu$m.
Black arrow points at the additional resonance peak in the spectra.
(b,c) 2D FFT of $M_z(x,t)$ maps at $\sigma = 0.2\,\mu$m and $\sigma = 0.5\,\mu$m, respectively.
}
\end{figure}

The increase of the pump spot width $\sigma$ at constant $\Delta K_u$ leads to narrowing of MSW spectra, as demonstrated in Fig.\ref{fig:FFT_vs_Sigma}(a).
The effect is a result of the decrease of the maximal wavevector $k_{max}$ if MSW excited at larger $\sigma$.
Therefore, in this case, the spectrum of exited MSW appears to be narrower with the suppressed high-frequency part (Fig.\ref{fig:FFT_vs_dK}(b)). 
The effect of the spectra narrowing is clearly seen in the reconstructed dispersions of MSW (Fig.\ref{fig:FFT_vs_Sigma}(b,c)).
Notably, for small values of $\sigma$ there are  additional resonance peaks appearing in the MSW spectra caused by DW-pump resonator.

\section{Conclusions}

We have shown the number of features of MSW optically excited by a femtosecond laser pulse in the vicinity of DW in the thin strip of a ferromagnet.
Firstly, the presence of DW makes it possible to excite MSW via the change of magnetic anisotropy induced by ultrafast laser heating in zero applied magnetic field, as the internal magnetic field of the stripe is non-uniform.
This contrasts with the case of the single domain state, when the excitation of MSW via such anisotropy change requires an applied magnetic field.
Secondly, focused optical pulse produces local changes in magnetic properties and, therefore, DW-pump spot system forms the resonator for MSW.
As a result of the interference of MSW propagating from the pump spot and reflected by the DW, the spectrum of the MSW packet outside the resonator possesses a complex structure with several resonance peaks.
The properties of the resonator depend on the pump parameters and position, enabling the adjustment of MSW spectrum.
For instance, the pump position allows tuning the frequencies and amplitudes of the peaks in the spectrum of MSW packets.
The increase of laser pulse fluence leads to the narrowing of MSW spectrum and the change of the ratio between the amplitudes of the resonance peaks in the spectrum.
Such tuning of the MSW packet is not available in traditional RF methods of SW excitation as a RF field does not vary the magnetic properties of SW guiding media.
Moreover, as the time scale of the anisotropy change with fs-laser pulses is of about 1\,ps, the presented SW resonator could be realized as an element of ultrafast optically reconfigurable magnonics \cite{VogelNPhys:2015, grundler_reconfigurable_2015}.

Finally, we note that the laser-induced anisotropy changes can also have different origins apart from ultrafast heating.
Thus, there are no principal restrictions on the sign and value of $\Delta K_u$, as well as on its temporal evolution.
Furthermore, the presented concept of MSW excitation and spectrum modification is expected to work near any magnetic non-uniformity inducing spatially varying stray and/or demagnetizing fields.
Such a non-uniformity can be induced by N\'eel or Bloch DW, magnetic skyrmion, bubble domain, impurities, etc.
Each type of non-uniformity presents an individual interest for the study as the internal structure, size, and topology affect its interaction with SW drastically \cite{Chang_SciRep2018ferromagneticDW_filter, Lan_SW_skyrmion_skew_PRB_2021}.
Importantly, the position and internal structure of magnetic non-uniformities, DW in particular, can be controlled by external magnetic and electric \cite{Pyatakov_magnetoelectricity_2017_JMMM} fields, spin-polarized currents \cite{Yamaguchi_Current-Driven_DW_PRL_2004}, and even by propagating SW \cite{Han_DWmotion_bySW_2009_APL, Chang_SciRep2018ferromagneticDW_filter, Dadoenkova_inelastic_SW_scattering_2019_PSSRRL, Yan_SWandDW_PRL_2011, Wang_DW_motion_via_SW_PRB_2012}, opening additional paths to tune the SW properties in magnonic devices.

\section*{Author statement}
\textbf{N.E. Khokhlov:} Methodology, Software, Writing - Review \& Editing, Supervision
\textbf{A.E. Khramova:} Conceptualization, Writing - Original Draft
\textbf{Ia.A. Filatov:} Formal analysis
\textbf{P.I. Gerevenkov:} Software, Visualization
\textbf{B.A. Klinskaya:} Visualization \textbf{A.M. Kalashnikova:} Conceptualization, Writing - Review \& Editing

\section*{acknowledgments}
The authors thank V.I. Belotelov for fruitful discussions.
N.E.Kh. and A.E.Kh. acknowledge financial support by the Russian Foundation for Basic Research (project No. 19-32-50128) and the "BASIS" Foundation (grants No. 19-1-3-42-1 and 18-2-6-202-1).

\section*{Competing interests}
The authors declare no competing interests.

\providecommand{\noopsort}[1]{}\providecommand{\singleletter}[1]{#1}%

\end{document}